\newcommand{\be}{\begin{equation}}
\newcommand{\ee}{\end{equation}}
\newcommand{\bea}{\begin{eqnarray}}
\newcommand{\eea}{\end{eqnarray}}

\def\rfr#1{eq.(\ref{#1})}

\def\Rfr#1{Eq.(\ref{#1})}

\def\eqi{\begin{equation}}
\def\eqf{\end{equation}}
\def\eqia{\begin{eqnarray}}
\def\eqfa{\end{eqnarray}}

\def\rp#1#2{{#1\over#2}}

\def\lb#1{\label{#1}}

\def\bm#1{{\mbox{\boldmath$#1$\unboldmath}}}

\documentclass[11pt]{article}

\usepackage{amsmath,amsthm,amscd,amssymb}
\usepackage{latexsym}
\usepackage{graphics,graphicx}

\begin{document}

\noindent{\bf \LARGE{The impact of the new Earth gravity models on
the measurement of
the Lense--Thirring effect}}
\\
\\
\\
 Lorenzo Iorio\\Dipartimento Interateneo di Fisica dell'
Universit${\rm \grave{a}}$ di Bari
\\Via Amendola 173, 70126\\Bari, Italy
\\
\\
Alberto Morea\\Dipartimento Interateneo di Fisica dell'
Universit${\rm \grave{a}}$ di Bari
\\Via Amendola 173, 70126\\Bari, Italy

\begin{abstract}
We examine how the new forthcoming Earth gravity models from the
CHAMP and, especially, GRACE missions could improve the
measurement of the general relativistic Lense--Thirring effect
according to the various kinds of observables which could be
adopted. In a very preliminary way, we use the first recently
released EIGEN2 CHAMP--only and GGM01C GRACE--based Earth gravity
models in order to assess the impact of the mismodelling in the
even zonal harmonic coefficients of geopotential which represents
one of the major sources of systematic errors in this kind of
measurement. However, discretion is advised on evaluating the
reliability of these results because the Earth gravity models used
here, especially EIGEN2, are still very preliminary and more
extensive calibration tests must be performed. According to the
GGM01C model, the systematic error due to the unmodelled even
zonal harmonics of geopotential amounts to 2$\%$ for the
combination of the nodes of LAGEOS and LAGEOS II and the perigee
of LAGEOS II used up to now by Ciufolini and coworkers in the
currently performed LAGEOS-LAGEOS II Lense-Thirring experiment,
and to 14$\%$ for a combination explicitly presented here which
involves the nodes only of LAGEOS and LAGEOS II.
\end{abstract}


\section{Introduction}
An interesting class of Post--Newtonian features is represented by
the orbital effects of order $\mathcal{O}(c^{-2})$ induced by the
linearized general relativistic gravitoelectromagnetic forces on
the motion of a test body freely falling in the gravitational
field of a central mass.

Among them, of great interest is the gravitomagnetic
Lense--Thirring effect or dragging of inertial frames \cite{Lense
and Thirring 1918, Ciufolini and Wheeler 1995} whose source is the
proper angular momentum \bm J of the central mass which acts as
source of the gravitational field. Its effect on the precessional
motion of the spins {\bm s} of four freely orbiting
superconducting gyroscopes should be tested, among other things,
by the important GP--B mission \cite{Everitt et al 2001} at a
claimed accuracy level of the order of 1$\%$ or better.

%, but such
%mission might be cancelled if it will fail any one of three tests
%recently drawn up by NASA managers \cite{Lawler 2003a; 2003b). However,
%at present,

Another possible way to measure such elusive relativistic effects
is the analysis of the laser--ranged data of some existing, or
proposed, geodetic satellites of LAGEOS--type as LAGEOS, LAGEOS II
\cite{Ciufolini 1996} and the proposed LAGEOS III--LARES
\cite{Ciufolini 1986, Iorio et al 2002, Iorio 2003b}. In this case
the whole orbit of the satellite is to be thought of as a giant
gyroscope whose longitude of the ascending node $\Omega$ and the
argument of perigee $\omega$ (In the original paper by Lense and
Thirring the longitude of the pericentre $\varpi=\Omega+\omega$ is
used instead of $\omega$) undergo the Lense--Thirring precessions
\begin{eqnarray}
\dot\Omega_{\rm LT} &=&\rp{2GJ}{c^2 a^3(1-e^2)^{\rp{3}{2}}},\\
\dot\omega_{\rm LT} &=&-\rp{6GJ\cos i}{c^2
a^3(1-e^2)^{\rp{3}{2}}},
\end{eqnarray} where $a,\ e$ and $i$ are the semimajor axis, the eccentricity and the inclination, respectively, of
the orbit and $G$ is the Newtonian gravitational constant. In
recent years first attempts would have yielded a measurement of
the Lense--Thirring dragging of the orbits of the existing LAGEOS
and LAGEOS II at a claimed accuracy of the order of $20\%-30\%$
\cite{Ciufolini et al 1998, Ciufolini 2002}. However, at present,
there are some scientists who propose different error budgets
\cite{Ries et al 2003}.
\section{The sources of error in the performed test}
The observable used in the tests reported in \cite{Ciufolini et al
1998, Ciufolini 2002} is the following linear combination of the
orbital residuals of the nodes of LAGEOS and LAGEOS II and the
perigee of LAGEOS II \cite{Ciufolini 1996}
\eqi\delta\dot\Omega^{\rm LAGEOS }+c_1\delta\dot\Omega^{\rm
LAGEOS\ II}+c_2\delta\dot\omega^{\rm LAGEOS\ II}\sim \mu_{\rm
LT}60.2,\lb{ciufform}\eqf where $c_1=0.304$, $c_2=-0.350$ and
$\mu_{\rm LT}$ is the solved--for least square parameter which is
0 in Newtonian mechanics and 1 in General Relativity. The
gravitomagnetic signature is a linear trend with a slope of 60.2
milliarcseconds per year (mas yr$^{-1}$ in the following).

The latest, 2002, measurement of the Lense--Thirring effect,
obtained by processing the LAGEOS and LAGEOS II data over a time
span of almost 8 years with the orbital processor GEODYN II of the
Goddard Space Flight Center, yields \cite{Ciufolini 2002}
\eqi\mu_{\rm LT}\sim 1\pm 0.02\pm\delta\mu_{\rm LT}^{\rm
systematic},\eqf where $\delta\mu_{\rm LT}^{\rm systematic}$
accounts for all the possible systematic errors due to the
mismodelling in the various competing classical forces of
gravitational and non--gravitational origin affecting the motion
of the LAGEOS satellites. In \cite{Ciufolini 2002} $\delta\mu_{\rm
LT}^{\rm systematic}$ is estimated to be of the order of
$20\%$--$30\%$.

The main source of gravitational errors is represented by the
aliasing classical secular precessions induced on the node and the
perigee of a near Earth satellite by the mismodelled even zonal
coefficients of the multipolar expansion of Earth gravitational
field: indeed, they mimic the genuine relativistic
trend\footnote{Another source of error which would plague an
attempted measurement of the Lense--Thirring effect with only one
orbital element would be the so called Lense--Thirring `imprint'.
It consists of the fact that in the solutions of the various Earth
gravity models General Relativity is assumed to be true, so that
the recovered $J_{l}$ are biased by this a priori assumption.
Then, any claimed measurement of the gravitomagnetic precessions
based, among other things, on such recovered values of the even
zonal harmonics would lack in full credibility and reliability. It
turns out that such sort of Lense--Thirring `imprint' is
concentrated, at least for the LAGEOS satellites, mainly in the
first two--three even zonal harmonics \cite{Ciufolini 1996} which
do affect the single orbital elements. }. \Rfr{ciufform} is
designed in order to cancel out the effects of the first two even
zonal harmonics of geopotential which induce mismodelled
precessions of the same order of magnitude, or even larger, than
the gravitomagnetic shifts, according to the Earth gravity model
EGM96 \cite{Lemoine et al 1998} (See Table
\ref{singolielementiegm96}). The evaluation of the impact of the
remaining uncancelled even zonal harmonics of higher degree on
\rfr{ciufform} is of the utmost importance. According to a
Root--Sum--Square calculation \cite{Iorio 2003a} with the full
covariance matrix of EGM96 up to degree $l=20$ it amounts to
almost
%\footnote{This result holds also if the calculations are
%the orbits of the LAGEOS satellites are insensitive to the even
%zonal harmonics of degree higher than $l=10$--$12$. See below for
%further explanations.}
13$\%$. However, according to the authors of \cite{Ries et al
2003}, it would not be entirely correct to automatically extend
the validity of the covariance matrix of EGM96, which is based on
a multi--year average that spans the 1970, 1980 and early 1990
decades, to any particular time span like that, e.g., of the
LAGEOS--LAGEOS II analysis which extends from the middle to the
end of the 1990 decade. Indeed, there would not be assurance that
the errors in the even zonal harmonics of the geopotential during
the time of the LAGEOS--LAGEOS II experiment remained correlated
exactly as in the EGM96 covariance matrix, in view of the various
secular, seasonal and stochastic variations that we know occur in
the terrestrial gravitational field and that have been neglected
in the EGM96 solution. Of course, the same would also hold for any
particular future time span of some years. If, consequently, the
diagonal part only of the covariance matrix of EGM96 is used, the
error due to geopotential, calculated in a Root--Sum--Square
fashion, i.e. by taking the square root of the sum of the squares
of the individual errors induced by the various even zonal
harmonics, amounts to almost\footnote{It is interesting to note
that, according to the diagonal part only of the covariance matrix
of the GRIM5--C1 Earth gravity model \cite{Gruber et al 2000}, the
RSS error due to the uncancelled even zonal harmonics amounts to
13.3$\%$. The GRIM5--S1 and GRIM5--C1 models represent the latest
solutions based on conventional satellite tracking data of the
pre--CHAMP and GRACE era. They are well tested and calibrated with
respect to other existing models \cite{Klokocnik et al 2002}.}
$45\%$ \cite{Iorio 2003a}. A really conservative upper bound of
the error due to geopotential is given by the sum of the absolute
values of the individual errors for the various even zonal
harmonics. For EGM96 it amounts to 83$\%$ (See Table
\ref{resultscomb}). Note that in the EGM96 solution (and in the
previous Earth gravity models like JGM3) the recovered even zonal
harmonics are highly correlated; in fact, it is likely that the
optimistic 13$\%$ result obtained with the full covariance matrix
is due to a lucky correlation between $J_6$ and $J_8$ \cite{Ries
et al 2003}. Then, in this case, the sum of the absolute values of
the individual errors should represent a truly realistic estimate
of the impact of the misomodelled even zonal harmonics of
geopotential.

Another important class of systematic errors is given by the
non--gravitational perturbations which affect especially the
perigee of LAGEOS II. For this subtle and intricate matter we
refer to \cite{Lucchesi 2001, Lucchesi 2002}. The main problem is
that it turned out that their interaction with the structure of
LAGEOS II changes in time due to unpredictable modifications in
the physical properties of the LAGEOS II surface (orbital
perturbations of radiative origin, e.g. the solar radiation
pressure and the Earth albedo) and in the evolution of the spin
dynamics of LAGEOS II (orbital perturbations of thermal origin
induced by the interaction of the electromagnetic radiation of
solar and terrestrial origin with the physical structure of the
satellites, in particular with their corner--cube
retroreflectors). Moreover, such tiny but insidious effects were
not entirely modelled in the GEODYN II software at the time of the
analysis of \cite{Ciufolini et al 1998}, so that it is not easy to
correctly and reliably assess their impact on the total error
budget of the measurement performed during that particular time
span. According to the evaluations in \cite{Lucchesi 2002}, the
systematic error due to the non--gravitational perturbations over
a time span of 7 years amounts to almost 28$\%$. However,
according to \cite{Ries et al 2003}, their impact on the
measurement of the Lense--Thirring effect with the nodes of LAGEOS
and LAGEOS II and the perigee of LAGEOS II is, in general, quite
difficult to be reliably assessed.
%Moreover, also
%the formal, standard statistical error can be notably influenced
%if, e.g., the effect of the direct solar radiation pressure
%effects on the perigee of LAGEOS II is not suitably accounted for
%in the data analysis.

So, by adding quadratically the gravitational and
non--gravitational errors of\footnote{The estimates obtained there
are based on levels of accuracy in knowing the non-gravitational
forces which do not coincide with those of the force models
included in GEODYN when the analysis of \cite{Ciufolini et al
1998} was performed.} \cite{Lucchesi 2002} we obtain for the
systematic uncertainty $\delta\mu_{\rm LT}^{\rm systematic}\sim
30\%$ if we assume a 13$\%$ error due to geopotential, and
$\delta\mu_{\rm LT}^{\rm systematic}\sim 54\%$ if we assume a
45$\%$ error due to geopotential. The sum of the absolute values
of the errors due to gepotential added quadratically with the
non--gravitational perturbations would yield a total systematic
error of $\delta\mu_{\rm LT}^{\rm systematic}\sim$ 87.6$\%$. It
must be noted that the latter estimate is rather similar to those
released in \cite{Ries et al 2003}. Moreover, it should be
considered that the perigee of LAGEOS II is also sensitive to the
eclipses effect on certain non--gravitational perturbations. Such
features are, generally, not accounted for in all such estimates.
An attempt can be found in \cite{Vespe 1999} in which the impact
of the eclipses on the effect of the direct solar radiation
pressure on the LAGEOS--LAGEOS II Lense--Thirring measurement has
been evaluated: it should amount to almost 10$\%$ over an
observational time span of 4 years.
\section{The opportunities offered by the new terrestrial gravity models}
From the previous considerations it could be argued that, in order
to have a rather precise and reliable estimate of the total
systematic error in the measurement of the Lense--Thirring effect
with the LAGEOS satellites it would be better to reduce the impact
of the geopotential in the error budget and/or discard the perigee
of LAGEOS II which is very difficult to handle and is a relevant
source of uncertainty due to its great sensitivity to many
non--gravitational perturbations.

The forthcoming more accurate Earth gravity models from the CHAMP
\cite{Pavlis 2000} and, especially, GRACE \cite{Ries  et al 2002}
missions, if the great expectations related to the latter will be
finally confirmed, will yield an opportunity to realize both these
goals, at least to a certain extent. In order to evaluate
quantitatively the opportunities offered by the new terrestrial
gravity models we have preliminarily used the recently released
EIGEN2 gravity model \cite{Reigber et al 2003}. It is a CHAMP-only
gravity field model derived from CHAMP GPS
satellite--to--satellite and accelerometer data out of the period
2000, July to December and 2002, September to December.  Although
higher degree and order terms are solved in EIGEN2, the solution
has full power only up to about degree/order 40 due to signal
attenuation in the satellite's altitude. Higher degree/order terms
are solvable applying regularization of the normal equation
system. However, in the case of the LAGEOS satellites it does not
pose problems because their nodes and perigees are sensitive to
just the first five--six even zonal harmonics\footnote{This means
that the error in the Lense--Thirring measurement due to the even
zonal harmonics of geopotential does not change any more if the
even zonal harmonic coefficients of degree higher than 10--12 are
neglected in the calculation.}. It is important to note that for
EIGEN2 it is likely that the released sigmas of the even zonal
harmonic coefficients, which are the formal errors, are rather
optimistic, at least for the low degree even zonal harmonics up to
$l=20$ \cite{Reigber et al 2003}.

In Table \ref{singolielementiegm96} we quote the errors in the
measurement of the Lense--Thirring effect with single orbital
elements of the LAGEOS satellites according to EGM96 up to degree
$l=70$ (See also Table II of \cite{Iorio 2003a}).
%%%%%%%%%%%%%%%%%%%%%%%%%%%%%%%%%%%%%%%%%%%%%%%%%%%%%%%
\begin{table}[ht!]
\caption{Systematic gravitational errors $\delta\mu_{\rm LT}^{\rm
systematic\ even\ zonals }$ in the measurement of the
Lense--Thirring effect with the nodes of the LAGEOS satellites and
the perigee of LAGEOS II only according to the EGM96 Earth gravity
model up to degree $l=70$. (C) denotes the full covariance matrix
while (D) refers to the diagonal part only used in a RSS way. A
pessimistic upper bound has been, instead, obtained from the sum
of the absolute values of the individual errors (SAV). In the
fifth column the impact of the mismodelling in $\dot J_2^{\rm
eff}$ over one year, according to \cite{Deleflie et al 2003}, is
quoted.  The effective coefficient $\dot J_2^{\rm eff}$ accounts
for the secular variations of the even zonal harmonics (see
below).} \label{singolielementiegm96}
\begin{tabular}{lllll}
\noalign{\hrule height 1.5pt}
LT (mas yr$^{-1})$ & percent error (C) & percent error (D) & percent error (SAV) & $\delta(\dot J_2^{\rm eff})$\\
\hline
$\dot\Omega^{\rm LAGEOS}_{\rm LT}$=30.7 & 50.3$\%$ & 199$\%$ & 341$\%$ & 8$\%$\\
$\dot\Omega^{\rm LAGEOS\ II}_{\rm LT}$=31.6 & 108$\%$ & 220 $\%$ & 382$\%$ & 14$\%$\\
$\dot\omega^{\rm
LAGEOS\ II }_{\rm LT}$ =-57.5 & 93$\%$ & 242$\%$ & 449$\%$ & 5.4$\%$\\
\noalign{\hrule height 1.5pt}
\end{tabular}
\end{table}
%%%%%%%%%%%%%%%%%%%%%%%%%%%%%%%%%%%%%%%%%%%%%%%%%%%%%%%
In Table \ref{singolielementieigen2} we quote the errors in the
measurement of the Lense--Thirring effect with single orbital
elements of the LAGEOS satellites according to EIGEN2
\footnote{The correlation matrix of EIGEN2 is downloadable from
http://op.gfz-potsdam.de/champ/results/ in the form of lower
triangular matrix. In it the recovered even zonal harmonics are
disentangled to a higher degree than in EGM96, so that a
Root--Sum--Square calculation with the variance matrix should be
adequate in reliably assessing the systematic error induced by the
mismodelled even zonal harmonics of geopotential. } up to degree
$l=70$.
%%%%%%%%%%%%%%%%%%%%%%%%%%%%%%%%%%%%%%%%%%%%%%%%%%%%%%%
\begin{table}[ht!]
\caption{Systematic gravitational errors $\delta\mu_{\rm LT}^{\rm
systematic\ even\ zonals }$ in the measurement of the
Lense--Thirring effect with the nodes of the LAGEOS satellites and
the perigee of LAGEOS II only according to the EIGEN2 Earth
gravity model up to degree $l=70$. (C) denotes the full covariance
matrix while (D) refers to the diagonal part only used in a RSS
way. A pessimistic upper bound has been, instead, obtained from
the sum of the absolute values of the individual errors (SAV). In
the fifth column the impact of the mismodelling in $\dot J_2^{\rm
eff}$ over one year, according to \cite{Deleflie et al 2003}, is
quoted. The effective coefficient $\dot J_2^{\rm eff}$ accounts
for the secular variations of the even zonal harmonics (see
below).} \label{singolielementieigen2}
\begin{tabular}{lllll}
\noalign{\hrule height 1.5pt}
LT (mas yr$^{-1})$ & percent error
(C) & percent error (D) & percent error (SAV) &
 $\delta(\dot J_2^{\rm eff})$\\
 \hline
 $\dot\Omega^{\rm LAGEOS}_{\rm LT}$=30.7 & 71.5$\%$ & 69$\%$ & 108$\%$ & 8$\%$\\
$\dot\Omega^{\rm LAGEOS\ II}_{\rm LT}$=31.6 & 107$\%$ & 107$\%$ & 144$\%$ & 14$\%$\\
$\dot\omega^{\rm
LAGEOS\ II }_{\rm LT}$ =-57.5 & 65$\%$ & 63$\%$ & 116$\%$ & 5.4$\%$\\
\noalign{\hrule height 1.5pt}
\end{tabular}
\end{table}
%%%%%%%%%%%%%%%%%%%%%%%%%%%%%%%%%%%%%%%%%%%%%%%%%%%%%%%
It can be noticed that, for EIGEN2, the results obtained with the
variance matrix in a Root--Sum--Square way are much more similar
to those obtained with the full covariance matrix than for EGM96;
this fact could be explained by noting that the even zonal
harmonics are better resolved and uncorrelated in EIGEN2 than in
EGM96 for which, instead, some favorable correlations may finally
yield the obtained results (See also \cite{Ries et al 2003}). The
simple sum of the absolute values of the individual errors for the
various degrees yields a pessimistic upper bound of the error due
to the bad knowledge of geopotential. However, Table
\ref{singolielementieigen2} clearly shows that the use of single
orbital elements of the LAGEOS satellites in order to measure the
Lense--Thirring effect is still unfeasible. Moreover, when a
single orbital element is analyzed, the effects of the secular
variation of the even zonal harmonics have to be considered as
well. It turns out that they can be accounted for by an effective
time rate \cite{Eanes and Bettadpur 1996} \eqi\dot J_2^{\rm
eff}\sim\dot J_2+0.371\dot J_4+0.079\dot J_6+0.006\dot
J_8-0.003\dot J_{10}...\eqf whose magnitude is of the order of
$(-2.6\pm 0.3)\times 10^{-11}$ yr$^{-1}$. Its impact on a possible
Lense--Thirring measurement is not negligible at all. It has been
evaluated, in a conservative way, by doubling the difference
between the maximum and minimum values of the adjusted $\dot
J_2^{\rm eff}$ for the longest arcs of Table 1 in \cite{Deleflie
et al 2003}, according to an approach followed in \cite{Lucchesi
2003}.

With regard to \rfr{ciufform}, it turns out that the systematic
error due to the even zonal harmonics of the geopotential,
according to the full covariance matrix of EIGEN2 up to degree
$l=70$, amounts to 7$\%$, while if the diagonal part\footnote{It
should be noted that the correlations represent the state of
processing of the about seven months of CHAMP data incorporated in
the EIGEN2 solution. No temporal variations in the zonal
coefficients were solved for, so no evolution of coefficients and
their correlations can be predicted directly from the solution. In
future it will be tried to resolve temporal variations from
solutions covering different data epochs (P. Schwintzer, private
communication). So, a possible conservative approach might consist
in using only the diagonal part of the covariance matrix. However,
the calibration of EIGEN2 errors should be extensively and
exhaustively checked.} only is adopted it becomes 9$\%$ (RSS
calculation). The sum of the absolute values yields an upper bound
of 16$\%$ (See Table \ref{resultscomb}). Of course, even if the
LAGEOS and LAGEOS II data had been reprocessed with the EIGEN2
model, the problems posed by the correct evaluation of the impact
of the non--gravitational perturbations on the perigee of LAGEOS
II would still persist,
%The
%total systematic error would still remain of the order of
%28$\%$--30$\%$,
unless significant improvements in the modeling of
the non--gravitational perturbations on the perigee of LAGEOS II
will occur.

A possible approach could be the use of linear combinations of
orbital residuals of the nodes and the perigees of the other
existing geodetic satellites of LAGEOS type like Starlette,
Stella, Ajisai, etc., so to cancel out as many even zonal
harmonics as possible. In \cite{Iorio 2002a, Iorio 2002b}, in
which the full covariance matrix of EGM96 up to degree $l=20$ has
been used, it turned out that, due to the lower altitude of the
other satellites to be employed, they are more sensitive than the
LAGEOS satellites to the even zonal harmonics of higher degree of
the geoptential and the combinations including their orbital
elements are not competitive with those including only the
LAGEOS--LAGEOS II elements. The following combination, which
includes the node of Ajisai, seemed to yield a slight improvement
in the systematic gravitational error \eqi\delta\dot\Omega^{\rm
LAGEOS}+c_1\delta\dot\Omega^{\rm LAGEOS \
II}+c_2\delta\dot\Omega^{\rm Ajisai}+c_3\delta\dot\omega^{\rm
LAGEOS \ II}\sim \mu_{\rm LT}61.2\lb{ajicomb},\eqf with
$c_1=0.443,\ c_2=-0.0275,\ c_3=-0.341$ Indeed, according to the
full covariance matrix of EGM96 up to degree $l=70$, it would
amount to 10.3$\%$. Note that it turns out that, with the
inclusion of Ajisai, the first ten even zonal harmonics have full
power in affecting the systematic error due to geopotential in the
Lense--Thirring measurement. If the correlations among the even
zonals are neglected, the variance matrix of EGM96, used in a RSS
calculation, yields a 64.4$\%$ error. The sum of the absolute
values of the individual errors yields an upper bound of 82$\%$.
By using the covariance matrix of EIGEN2 up to degree $l=70$ the
systematic gravitational error raises to 13.4$\%$ (13.6$\%$ with
the diagonal part only of the covariance matrix of EIGEN2 up to
degree $l=70$. RSS calculation.). The sum of the absolute values
of the individual errors yields an upper bound of 16$\%$ (See
Table \ref{resultscomb}). Since the non--gravitational part of the
error budget of \rfr{ajicomb} is almost similar to that of
\rfr{ciufform}, as it can be inferred from the magnitude of the
coefficients of \rfr{ajicomb} and \rfr{ciufform} which weigh the
various orbital elements, it is obvious that \rfr{ajicomb} would
not represent any substatntial improvement with respect to the
LAGEOS--LAGEOS II observable of \rfr{ciufform}.
\subsection{A new nodes--only combination}
A different approach could be followed by taking the drastic
decision of canceling out only the first even zonal harmonic of
geopotential by discarding at all the perigee of LAGEOS II. The
hope is that the resulting gravitational error is reasonably small
so to get a net gain in the error budget thanks to the fact that
the nodes of LAGEOS and LAGEOS II exhibit a very good behavior
with respect to the non--gravitational perturbations. Indeed, they
are far less sensitive to their tricky features than the perigee
of LAGEOS II. Moreover, they can be easily and accurately
measured, so that also the formal, statistical error should be
reduced. A possible combination is \eqi\delta\dot\Omega^{\rm
LAGEOS }+c_1\delta\dot\Omega^{\rm LAGEOS\ II}\sim \mu_{\rm
LT}48.2,\lb{iorform}\eqf where $c_1= 0.546$. A similar approach is
proposed in \cite{Ries et al 2002}, although without quantitative
details. According to the full covariance matrix of EIGEN2 up to
degree $l=70$, the systematic error due to the even zonal
harmonics from $l=4$ to $l=70$ amounts to 8.5 mas yr$^{-1}$
yielding a 17.8$\%$ percent error, while if the diagonal part only
is adopted it becomes 22$\%$ (RSS calculation). EGM96 would not
allow to adopt \rfr{iorform} because its full covariance matrix up
to degree $l=70$ yields an error of 47.8$\%$ while the error
according to its diagonal part only amounts even to\footnote{It
reduces to 60$\%$ according to the diagonal part only of the
covariance matrix of the GRIM5--C1 model (RSS calculation).}
104$\%$ (RSS calculation: see Table \ref{resultscomb}). Note also
that \rfr{iorform} preserves one of the most important features of
the other combinations of orbital residuals: indeed, it allows to
cancel out the very insidious 18.6-year tidal perturbation which
is a $l=2,\ m=0$ constituent with a period of 18.6 years due to
the Moon's node and nominal amplitudes of the order of 10$^3$ mas
on the nodes of LAGEOS and LAGEOS II \cite{Iorio 2001}. Moreover,
also the secular variations of the even zonal harmonic
coefficients of geopotential do not affect the proposed
combination: indeed, \rfr{iorform} is designed in order to cancel
out just all the effects of the first even zonal harmonic
coefficient. On the other hand, the impact of the
non--gravitational perturbations on \rfr{iorform} over a time span
of, say, 7 years can be quantified in just 0.1 mas yr$^{-1}$,
yielding a 0.3$\%$ percent error. The results of Tables 2 and 3 of
\cite{Iorio et al 2002} have been applied to \rfr{iorform} by
adding in quadrature the various mismodelled perturbing effects
for such combination of orbital elements. To them a 20$\%$
mismodelling in the Yarkovsky--Rubincam and Yarkovsky--Schach
effects and Earth's albedo and a 0.5$\%$ mismodelling in the
direct solar radiation pressure have been applied. It is also
important to notice that, thanks to the fact that the periods of
many gravitational and non--gravitational time--dependent
perturbations acting on the nodes of the LAGEOS satellites are
rather short, a reanalysis of the LAGEOS and LAGEOS II data over
just a few years could be performed. So, with a little
time--consuming reanalysis of the nodes only of the existing
LAGEOS and LAGEOS II satellites with the EIGEN2 data it would at
once be possible to obtain a more accurate and reliable
measurement of the Lense--Thirring effect, avoiding the problem of
the uncertainties related to the use of the perigee of LAGEOS II.
Moreover, it should be noted that the forthcoming, more accurate
and robust solutions of the terrestrial gravity fields including
the data from both CHAMP and GRACE should yield better results for
the systematic error due to the geopotential. Of course, in order
to push the gravitational error at the level of a few percent a
new LAGEOS--like satellite as the proposed LARES should at least
be used \cite{Iorio et al 2002, Iorio 2003b}.
%%%%%%%%%%%%%%%%%%%%%%%%%%%%%%%%%%%%%%%%%%%%%%%%%%%%%%%
\begin{table}[ht!]
\caption{Systematic gravitational errors $\delta\mu_{\rm LT}^{\rm
systematic\ even\ zonals }$ of various combinations of orbital
residuals according to EGM96 and EIGEN2 Earth gravity models up to
degree $l=70$. Ob. refers to the combination of orbital residuals
adopted. $\mathcal{C}$ refers to the Ciufolini's combination of
\rfr{ciufform}, $\mathcal{A}$ refers to the combination of
\rfr{ajicomb} which includes the node of Ajisai and $\mathcal{I}$
refers to the nodes--only combination of \rfr{iorform} presented
here. (C) denotes the use of the full covariance matrix while (D)
refers to the diagonal part only (RSS calculation). (SAV) denotes
the upper bound obtained from the sum of the absolute values of
the individual errors. } \label{resultscomb}
\begin{tabular}{lllllll}
\noalign{\hrule height 1.5pt}
 Ob. & EGM96 (C) & EGM96 (D) &  EGM96 (SAV)& EIG2 (C) & EIG2 (D) & EIG2 (SAV)\\
\hline $\mathcal{C}$ & 12.9$\%$ & 45$\%$ & 83$\%$& 7$\%$ & 9$\%$ & 16$\%$\\
$\mathcal{A}$ & 10.3$\%$ & 64.4$\%$ & 152$\%$ & 13.4$\%$ & 12.8$\%$ & 31.8$\%$\\
$\mathcal{I}$  &  48$\%$ & 104$\%$ & 177$\%$& 17.8$\%$ & 22$\%$ & 37$\%$\\
\noalign{\hrule height 1.5pt}
\end{tabular}
\end{table}
%%%%%%%%%%%%%%%%%%%%%%%%%%%%%%%%%%%%%%%%%%%%%%%%%%%%%%%
\subsection{First promising results from GRACE}
Very recently the first preliminary Earth gravity models including
some data from GRACE have been released; among them the GGM01C
model\footnote{It can be retrieved on the WEB at
http://www.csr.utexas.edu/grace/gravity/}, which combines the
Center for Space Research (CSR) TEG-4 model\footnote{The
GRACE--only GGM01S model was combined with the TEG-4 information
equations (created from historical multi--satellite tracking data;
surface gravity data and altimetric sea surface heights) to
produce the preliminary gravity model GGM01C.} \cite{Tapley et al
2000} with data from GRACE, seems to be very promising for our
purposes. Indeed, the released sigmas are not the mere formal
errors but are approximately calibrated. See Table
\ref{singolielementiGGM01C} for the effect on the single elements;
the improvements with respect to Table \ref{singolielementiegm96}
and Table \ref{singolielementieigen2} are evident, although not
yet sufficiently good in order to allow for a rather accurate
measurement of the Lense--Thirring effect by means of only one
orbital element.

The error due to geopotential in the combination of
\rfr{ciufform}, evaluated by using the variance matrix only in a
Root--Sum--Square fashion, amounts to 2.2$\%$ (with an upper bound
of 3.1$\%$ obtained from the sum of the absolute values of the
individual terms). Instead, the combination of \rfr{iorform} would
be affected at almost 14$\%$ level (RSS calculation), with an
upper bound of almost 18$\%$ from the sum of the absolute values
of the single errors. According also to GGM01C, the combination of
\rfr{ajicomb} seems to be not particularly competitive with
respect to that of \rfr{ciufform}. Indeed, the RSS error amounts
to 0.8$\%$, while the upper bound due to the sum of the absolute
values of the individual errors is of the order of 2$\%$. See
Table \ref{singolielementiGGM01C} also for the combinations of
orbital elements. Note that also for GGM01C the covariance matrix
is almost diagonal, so that the Root--Sum--Square calculations
should yield a realistic evaluation of the systematic error due to
the even zonal harmonics of geopotential.
%%%%%%%%%%%%%%%%%%%%%%%%%%%%%%%%%%%%%%%%%%%%%%%%%%%%%%%
\begin{table}[ht!]
\caption{Systematic gravitational errors $\delta\mu_{\rm LT}^{\rm
systematic\ even\ zonals }$ in the measurement of the
Lense--Thirring effect with the nodes of the LAGEOS satellites and
the perigee of LAGEOS II only and with some combinations according
to the GGM01C Earth gravity model up to degree $l=70$. The sigmas
of the even zonal coefficients of this solution are not the mere
formal errors but are approximately calibrated. $\mathcal{C}$
refers to the Ciufolini's combination of \rfr{ciufform},
$\mathcal{A}$ refers to the combination of \rfr{ajicomb} which
includes the node of Ajisai and $\mathcal{I}$ refers to the
nodes--only combination of \rfr{iorform} presented here.
$\mathcal{M}$ refers to the multi-satellite combination of
\rfr{multico}. (D) refers to the diagonal part only used in a RSS
way. A pessimistic upper bound has been, instead, obtained from
the sum of the absolute values of the individual errors (SAV). In
the fourth column the impact of the mismodelling in $\dot J_2^{\rm
eff}$ over one year, according to \cite{Deleflie et al 2003}, is
quoted.} \label{singolielementiGGM01C}
\begin{tabular}{llll}
\noalign{\hrule height 1.5pt}
LT (mas yr$^{-1})$ & percent error (D) & percent error (SAV) & $\delta(\dot J_2^{\rm eff})$\\
\hline $\dot\Omega^{\rm LAGEOS}_{\rm LT}$=30.7 & 44$\%$ & 66$\%$ & 8$\%$\\
$\dot\Omega^{\rm LAGEOS\ II}_{\rm LT}$=31.6 & 64$\%$ & 79$\%$ & 14$\%$\\
$\dot\omega^{\rm
LAGEOS\ II }_{\rm LT}$ =-57.5& 43$\%$  & 65$\%$ & 5.4$\%$\\
$\mathcal{C}=60.2$ & 2$\%$   & 3$\%$   & -\\
$\mathcal{A}=61.2$ & 0.8$\%$ & 1.8$\%$ & -\\
$\mathcal{I}=48.2$ & 14$\%$  & 18$\%$  & -\\
$\mathcal{M}=57.4$ & 37$\%$  & 123$\%$ & -\\  \noalign{\hrule
height 1.5pt}
\end{tabular}
\end{table}
%%%%%%%%%%%%%%%%%%%%%%%%%%%%%%%%%%%%%%%%%%%%%%%%%%%%%%%
It may be interesting to consider the following combination
\eqi\delta\dot\Omega^{\rm LAGEOS}+c_1\delta\dot\Omega^{\rm LAGEOS\
II}+c_2\delta\dot\Omega^{\rm Ajisai}+c_3\delta\dot\Omega^{\rm
Starlette}+c_4\delta\dot\Omega^{\rm Stella}\sim \mu_{\rm
LT}57.4,\lb{multico}\eqf with $c_1= 4.174,\ c_2=-2.705,\
c_3=1.508,\ c_4=-0.048$. According to a Root--Sum--Square
calculation with the variance matrix of GGM01C up to\footnote{It
has been checked that the error due to the even zonal harmonics
remains stable if other even zonal harmonics are added to the
calculation. Moreover, it turns also out that, up to $l=42$ there
are no appreciable fluctuations in the calculated classical
secular precessions.} $l=42$ the impact of the remaining even
zonal harmonics of degree $l\geq 10$ amounts to 21.6 mas yr$^{-1}$
which yields a $37.6\%$ percent error in the measurement of the
Lense--Thirring effect with \rfr{multico}. The upper bound due to
the sum of the absolute values of the individual errors amounts to
123$\%$. If the future GRACE--based gravity solutions will improve
the high degree ($J_{10},\ J_{12},\ J_{14},...$) even zonal
harmonics more than the low degree ($J_2,\ J_4,\ J_6,\ J_8$) ones,
the combination of \rfr{multico} could deserve some interest in
alternative to that of \rfr{iorform}.
\section{Conclusions}
In this paper we have used, in a very preliminary way, the data
from the recently released EIGEN2 Earth gravity model, including
six months of CHAMP data, in order to reassess the systematic
error due to the even zonal harmonics of the geopotential in the
LAGEOS--LAGEOS II Lense--Thirring experiment. The main results are
summarized in Table \ref{resultscomb} for EGM96 and EIGEN2 and
Table \ref{singolielementiGGM01C} for GGM01C which includes the
first data from GRACE.  It turned out that, by neglecting the
correlations between the various harmonics, such kind of error
changes from 45$\%$ (or, perhaps more realistically, 83$\%$) of
EGM96 to 9$\%$ of EIGEN2. Since the correct evaluation of the
error budget of such experiment is plagued by the uncertainties
due to the impact of the non--gravitational perturbations on the
perigee of LAGEOS II,
 we have considered an
observable including only the nodes of LAGEOS and LAGEOS II. It
turns out that the systematic error due to the even zonal
harmonics of the geopotential, according to EIGEN2 and neglecting
the correlations between the various harmonics, amounts to 22$\%$.
However, such an observable is almost insensitive to the
non--gravitational perturbations which would enter the error
budget at a level lower than 1$\%$.

It must be emphasized that the EIGEN2 solution is very preliminary
and exhaustive tests should be conducted in order to assess
reliably the calibration of the claimed errors, especially in the
lower degree even zonal harmonics to which the orbits of the
LAGEOS satellites are particularly sensitive. If and when more
robust and confident solutions for the terrestrial gravitational
field will be hopefully available, especially from GRACE, the
proposed observable based on the nodes of the two LAGEOS
satellites only could represent a good opportunity for measuring
the Lense--Thirring effect in an efficient, fast and reliable way.
The first results obtained with the very preliminary GGM01C model,
which includes the first data from GRACE and for which the
tentatively calibrated errors are available, point toward this
direction.
%%%%%%%%%%%%%%%%%%%%%%%%%%%%%%%%%%%%%%%%%%%%%%%%%%%%%%%%%%%%%%%%%%%%%%%%%%%%
\section*{Acknowledgments}
We are grateful to L. Guerriero for his support while at Bari. L.
I. gratefully thanks F. Vespe for his kind suggestions and
fruitful discussions.
%-----------------------------------------

\end{document}